# Analysis of genetic differences between psychiatric disorders: exploring pathways and cell-types/tissues involved and ability to differentiate the disorders by polygenic scores

Shitao RAO[1,8,9]*, Liangying YIN[1]*, Yong XIANG[1], Hon-Cheong SO[1-7]^

[1]School of Biomedical Sciences, The Chinese University of Hong Kong, Shatin, Hong Kong
[2]KIZ-CUHK Joint Laboratory of Bioresources and Molecular Research of Common Diseases, Kunming Institute of Zoology and The Chinese University of Hong Kong, China
[3]CUHK Shenzhen Research Institute, Shenzhen, China
[4]Department of Psychiatry, The Chinese University of Hong Kong, Hong Kong
[5]Margaret K.L. Cheung Research Centre for Management of Parkinsonism, The Chinese University of Hong Kong, Shatin, Hong Kong
[6] Brain and Mind Institute, The Chinese University of Hong Kong, Shatin, Hong Kong
[7]Hong Kong Branch of the Chinese Academy of Sciences (CAS) Center for Excellence in Animal Evolution and Genetics, The Chinese University of Hong Kong, Shatin, Hong Kong
[8]Department of Bioinformatics, Fujian Key Laboratory of Medical Bioinformatics, School of Medical Technology and Engineering, Fujian Medical University, Fuzhou, China;
[9]Key Laboratory of Ministry of Education for Gastrointestinal Cancer, School of Basic Medical Sciences, Fujian Medical University, Fuzhou, China

*These authors contributed equally to this work
^**Correspondence to: Hon-Cheong So**, Lo Kwee-Seong Integrated Biomedical Sciences Building, The Chinese University of Hong Kong, Shatin, Hong Kong. Tel: +852 3943 9255; E-mail: hcso@cuhk.edu.hk

**Abstract**

Although displaying genetic correlations, psychiatric disorders are clinically defined as categorical entities as they each have distinguishing clinical features and may involve different treatments. Identifying *differential* genetic variations between these disorders may reveal how the disorders differ biologically and help to guide more personalized treatment.

Here we presented a comprehensive analysis to identify genetic markers *differentially* associated with various psychiatric disorders/traits based on GWAS summary statistics, covering 18 psychiatric traits/disorders and 26 comparisons. We also conducted comprehensive analysis to unravel the genes, pathways and SNP functional categories involved, and the cell types and tissues implicated. We also assessed how well one could distinguish between psychiatric disorders by polygenic risk scores(PRS).

SNP-based heritabilities ($h^2_{SNP}$) were significantly larger than zero for most comparisons. Based on current GWAS data, PRS have mostly modest power to distinguish between psychiatric disorders. For example, we estimated that AUC for distinguishing schizophrenia from major depressive disorder(MDD), bipolar disorder(BPD) from MDD and schizophrenia from BPD were 0.694, 0.602 and 0.618 respectively, while the maximum AUC(based on $h^2_{SNP}$) were 0.763, 0.749 and 0.726 respectively. We also uncovered differences in each pair of studied traits in terms of their differences in genetic correlation with comorbid traits. For example, clinically-defined MDD appeared to more strongly genetically correlated with other psychiatric disorders and heart disease, when compared to non-clinically-defined depression in UK Biobank.

Our findings highlight genetic differences between psychiatric disorders and the mechanisms involved. PRS may help differential diagnosis of selected psychiatric disorders in the future with larger GWAS samples.

**Introduction**

Psychiatric disorders are common and more than one-third of the population suffer from at least one kind of disorder in their life[1]. Psychiatric disorders also rank among the top in terms of total disability-adjusted life years (DALYs)[2] lost. Recent analyses based on genome-wide association studies (GWAS) have suggested a moderate to high genetic correlation between many psychiatric disorders[3, 4]. On the other hand, although displaying strong genetic correlations, these disorders are clinically defined as independent categorical entities as they each have distinguishing clinical symptoms and often require different treatments[5]. Identifying *differential* genetic variations between these disorders may shed light on how the disorders differ biologically and help to guide more personalized treatment in the future. Another potential clinical application is that genetic markers may help differential diagnosis (DDx) of related disorders. For example, a patient who presents with depression for the first episode may actually be having bipolar disorder (BPD). It is often difficult to distinguish the two diagnoses by clinical features alone at the first presentation, but their treatments differ in important ways. If genetic information can help differentiate BPD from unipolar depression, it will enable more appropriate treatments to be given at an earlier stage of illness.

Most genetic studies to date have focused on identifying shared loci or genetic overlap between psychiatric disorders[6]. An effort to explore genetic architecture differences between BPD and SCZ was made by a recent study[7, 8]. They first compared 9,252 BPD cases to 7,129 SCZ cases but did not find any SNPs reaching genome-wide significance[7]; however, polygenic risk score(PRS) analysis showed that the score significantly differed between SCZ and BPD patients, indicating that differences between the two disorders have a genetic basis. More recently, they conducted an association analysis with a larger sample size (23,585 SCZ cases and 15,270 BPD cases) and identified two genome-wide significant SNPs[8]. However, the above analyses require individual genotyping data, which might be difficult to access due to privacy concerns. In addition, many of the largest GWAS analyses were conducted by meta-analyses and typically only summary statistics are available.

Here we presented a comprehensive analysis to identify *differential* genetic markers covering 18 psychiatric disorders/traits and 26 comparisons, based on GWAS summary statistics. The analytic framework was successfully validated by simulation studies before applications. Our results based on GWAS summary data showed almost perfect genetic correlation with those obtained via comparing BPD and SCZ individual genotyping data[8]($r_g$ =1.054, se=0.025), suggesting that our approach is reliable and resembles results from individual-level data analysis.

Importantly, we also conducted in-depth analysis to reveal the genes and pathways involved, and which cell types and tissues were the most relevant in differentiating the disorders. We also uncovered differences in each pair of disorders in terms of how they are genetically related to different sets of comorbidities. Another novel contribution is that we assessed how well we could distinguish two psychiatric disorders (e.g., major depressive disorder[MDD] vs BPD) using PRS from existing GWAS data, as well as the maximum discriminating ability from all GWAS-panel variants. This may be clinically relevant in the future given the lack of biomarkers for differential diagnosis(DDx) of psychiatric disorders.

# Methods

For details of methods and samples, please also refer to the Supplementary Text.

## GWAS summary statistics

A set of GWAS summary statistics for 18 psychiatric disorders/traits were included (Table 2) which were obtained from several public databases, for example the Psychiatric Genomics Consortium(PGC, https://www.med.unc.edu/pgc/), the Complex Trait Genetics lab (CTG lab, https://ctg.cncr.nl/) and the UK Biobank (UKBB; http://www.nealelab.is/uk-biobank). Details of the datasets are given in Table 2 and references therein.

We included a total of 10 psychiatric disorders in our analysis, including MDD, post-traumatic stress disorder (PTSD), eating disorder (ED), schizophrenia (SCZ), bipolar disorder (BPD), autistic spectrum disorder (ASD), attention deficit/hyperactivity disorder (ADHD), anxiety disorder, obsessive-compulsive disorder (OCD) and alcohol dependence. In principle, we wish to select a wide range of psychiatric disorders covering mood, psychotic, neurotic/stress-related disorders and disorders related psychoactive substance use. Besides, we also included three other depression-related phenotypes to be compared against MDD from PGC[9]. These 3 phenotypes were based on the UKBB sample, including longest period of feeling low/depressed, seen doctor(GP) for nerves, anxiety, tension or depression (to represent self-reported non-specific depression/low mood), and probable recurrent major depression (severe). The latter was derived from several questions based on Smith et al. [10]. In addition to the above, we also included ever used cannabis, insomnia, suicide attempts (SA), neuroticism and psychotic experience in our analysis as they are closely related to many psychiatric disorders. For details on the choice of phenotypes, please refer to the Supplementary Text.

## Identification of differential genetic markers

We present an analytic approach capable of unravelling the genetic differences between a pair of disorders/traits, relying only on GWAS summary statistics. The method also allows overlap in study samples. In essence, we are 'mimicking' a case–control GWAS in which the cases are subjects affected with one disorder and controls affected with the other disorder.

Suppose $T_1$ and $T_2$ are two binary traits under study. Let $S$ be a biallelic SNP, coded as 0, 1 or 2. For simplicity, we first assume this is a prospective study of a population-based sample. Based on the principles of logistic regression, we have

$$\log\left(\frac{P(T_1=1)}{P(T_1=0)}\right) = \log(\frac{p_1}{1-p_1}) = \beta_{01} + \beta_{11}S + \varepsilon_1 \qquad 2.1$$

$$\log\left(\frac{P(T_2=1)}{P(T_2=0)}\right) = \log(\frac{p_2}{1-p_2}) = \beta_{02} + \beta_{12}S + \varepsilon_2 \qquad 2.2$$

$$\log\left(\frac{P(T_1=1)}{P(T_2=1)}\right) = \log(\frac{p_3}{1-p_3}) = \beta_{03} + \beta_{13}S + \varepsilon_3 \qquad 2.2$$

where $p_1 = P(T_1 = 1)$ and $p_2 = P(T_2 = 1)$ denote the probability of the corresponding traits in the collected dataset; $\varepsilon_i (i = 1,2,3)$ indicates the error term for corresponding regression model. Based on the definition of odds ratio (OR), for traits $T_1$ and $T_2$, we have:

$$OR(T_1 \text{ vs } ctrl) = e^{\beta_{11}} = \frac{\Pr(T_1=1|S=s+1,covariates)}{\Pr(T_1=0|S=s+1,covariates)} \Big/ \frac{\Pr(T_1=1|S=s,covariates)}{\Pr(T_1=0|S=s,covariates)} \qquad 2.3$$

$$OR(T_2 \text{ vs } ctrl) = e^{\beta_{12}} = \frac{\Pr(T_2=1|S=s+1,covariates)}{\Pr(T_2=0|S=s+1,covariates)} \Big/ \frac{\Pr(T_2=1|S=s,covariates)}{\Pr(T_2=0|S=s,covariates)} \qquad 2.4$$

Suppose the controls for the two studies come from the same population. In this regard, $\Pr(T_1 = 0|S = s + 1, covariates)$ and $\Pr(T_1 = 0|S = s, covariates)$ are approximately the same as $\Pr(T_2 = 0|S = s + 1, covariates)$ and $\Pr(T_2 = 0|S = s, covariates)$ respectively. Thus, the odds ratio (OR) for differential association between the two diseases can be given as:

$$OR(T1 \text{ vs } T2) = e^{\beta_{13}} = \frac{\Pr(T_1=1|S=s+1,covariates)}{\Pr(T_2=1|S=s+1,covariates)} \Big/ \frac{\Pr(T_1=1|S=s,covariates)}{\Pr(T_2=1|S=s,covariates)}$$

$$\approx$$

$$\left(\frac{\Pr(T_1=1|S=s+1,covariates)}{\Pr(T_1=0|S=s+1,covariates)} \Big/ \frac{\Pr(T_1=1|S=s,covariates)}{\Pr(T_1=0|S=s,covariates)}\right) \div \qquad 2.5$$

$$\left(\frac{\Pr(T_2=1|S=s+1,covariates)}{\Pr(T_2=0|S=s+1,covariates)} \Big/ \frac{\Pr(T_2=1|S=s,covariates)}{\Pr(T_2=0|S=s,covariates)}\right)$$

$$= e^{\beta_{11}-\beta_{12}}$$

In other words, the effect size of differential association (i.e. trait 1 as case and trait 2 as control) can be derived from the difference of effect sizes of the respective traits. The variance of $\beta_{13}$ can be expressed as:

$$Var(\beta_{13}) = Var(\beta_{11} - \beta_{12}) = Var(\beta_{11}) + Var(\beta_{12}) - 2Cov(\beta_{11}, \beta_{12}) \qquad 2.6$$

$Cov(\beta_{11}, \beta_{12})$ depends on the actual overlap between the samples and the correlation between the 2 phenotypes. It can be derived from multiplying the SEs of the two coefficients with the intercept from cross-trait LD score regression (LDSC) (see equation 6 in[11]). Note that the above derivations only require the regression coefficients (beta), which is the same under a prospective (population-based) or a retrospective design (case-control design where cases may be over- or under-sampled)[12].

Two traits(neuroticism and longest period of feeling depressed/low) were continuous traits. To be consistent with other comparisons which all involves binary traits/disorders, we considered the summary statistics of a corresponding case-control study in which subjects at *top 20%* of the outcome are considered as 'cases'. The method for deriving binary-trait summary statistics was described in[13]. After computing the differential genetic associations, to further protect against population stratification, we performed genomic control following[14] (i.e. genomic inflation factor was based on LDSC result).

To further check the validity of our approach, we also computed genetic correlation of the results from a GWAS of *BPD vs SCZ* from our analytic method against those obtained by comparing the two disorders directly using *individual* genotype data, reported in ref[8].

**Functional annotations of identified differential genetic markers/Gene mapping**

The differential genetic variants identified were further explored for their biological functions using FUMA (https://fuma.ctglab.nl/)[15]. SNPs were mapped to genes in FUMA using three different strategies including mapping by position, expression quantitative trait loci (eQTL), and chromatin interactions (CI). Details are given in The Supplementary Text.

**Genome-wide gene-based association study (GWGAS) and tissue/cell-type enrichment analysis**

*P*-values from SNP-based analysis were utilized for GWGAS analysis in MAGMA[16]. The biological functions of GWGAS-significant genes were further investigated *via* tissue and cell-type expression enrichment analysis using MAGMA[16] and LDSC[14]

**SNP-based heritability and genetic correlation with related traits**

A number of previous studies have shown that common genetic variants as a whole contribute significantly to the susceptibility of individual psychiatric disorders, such as schizophrenia and major depression[17-19]. Building on previous studies, here we ask a slightly different question. We wish to know whether (and to what extent) common variants as a whole would contribute to the *difference* in susceptibility to different psychiatric disorders. Intuitively, for instance, both SCZ and BPD are highly heritable, however to what extent do common variants determine why someone may develop SCZ instead of BPD (or vice versa)?

To answer the above question, SNP-based heritability ($h^2_{snp}$) of differential genetic associations was estimated by LDSC and SumHer [20]. We also conducted 'partitioned heritability' analysis to identify enriched functional categories[21]. Heritability explained is connected to the predictive power of variants[22]. In this regard, we also estimated the *maximum* ability to differentiate the disorders that can be achieved if all variants on the GWAS panel are accounted for. We followed ref[22] to compute different predictive indices and graphs. Briefly, we computed the AUC under ROC curve, proportion of cases explained by those at the top *k*% of predicted risk, variance of predicted risk and the absolute risk at different percentiles. The graphs which were used to visualize predictive performance included the ROC curve, predictiveness curve and the probability and cumulative density function of predicted risks. This analysis was performed on selected psychiatric disorders (see Table 5) and clinical symptoms (psychotic experience) for which differential diagnosis is considered more clinically relevant.

Genetic correlations ($r_g$) between the differential genetic variations and 42 potentially related phenotypes were calculated using LDSC (http://ldsc.broadinstitute.org/centers/)[23]. Generally, $r_g$ reflects how much the non-shared or unique genetic component of the 1st disorder is genetically correlated with a specific trait, when compared to the 2nd disorder in the pair.

**Potential ability of polygenic risk scores (PRS) from existing GWAS data to differentiate disorders**

We performed another analysis to evaluate the ability of polygenic risk scores (PRS) from *existing* GWAS data to distinguish psychiatric disorders. The PRS was based on a case-control study of the corresponding disorders (disorder A as 'case' and disorder B as 'control'). Note that unlike above, we are *not* focusing on the *maximum* predictive power achievable from all common variants.

An empirical Bayes approach has been proposed to recover the underlying effect sizes and could be used to forecast predictive ability of PRS, based on summary statistics alone[24]. The method has been verified in simulations and real data applications[24]. Eighteen subsets of genetic variants based on a series of *P*-value thresholds($10^{-5}, 10^{-4}, 5 \times 10^{-4}, 10^{-3}, 5 \times 10^{-3}, 0.01, 0.03, 0.05, 0.1, 0.2, 0.3, 0.4, 0.5, 0.6, 0.7, 0.8, 0.9$, and 1) were used to

construct PRS. Note that here we only employed summary statistics to estimate the potential discriminatory ability of PRS; no individual-level genotype data was used.

**Simulation**

To verify the validity of our proposed method, we simulated different sets of genotype-phenotype data assuming 300 biallelic SNPs($N_{snp}$=300) and two disorders. Since the proposed framework is a SNP-based analysis, the number of simulated SNPs will not affect the validity of our simulation. Allele frequency for each simulated SNP was randomly generated from a uniform distribution within [0.05, 0.95]. The number of subjects with each disorder (*ncases*) was set to [10000, 20000, 50000, 100000] with a disease prevalence (*K*) of 10%. Here, *ncases* denotes the expected number of cases in the whole simulated population cohort. Given the disease prevalence, the whole simulated population cohort (*ntotal*) has a sample size of $ntotal = \frac{ncases}{K}$. The total SNP-based heritability ($h^2_{snp}$) for each trait was set at 0.2 to 0.4, distributed across all SNPs.

From the simulated population cohort, we simulated two case-control studies with traits A and B as the outcome respectively. The objective is to simulate GWAS summary statistics which are used as input for our methodology. Suppose the number of cases for traits A and B in the simulated population cohorts are respectively $N_A$ and $N_B$, and $N = max(N_A, N_B)$. To construct the GWAS summary statistics, for trait A, we picked $N_A$ cases and $2N - N_A$ controls from the population. For trait B, we picked $N_B$ cases and $2N - N_B$ controls from the population. We conducted two sets of analysis. In the 1st set, we only considered cases without comorbid disorders, i.e. all cases identified as having trait A but not trait B were selected as cases ($N_{A\_only}$) and compared to population controls. Similarly, we conducted GWAS on $N_{B\_only}$ cases against population controls. This will mimic the situation for disorders that are not diagnosed together like MDD/BPD or SCZ/BPD, or if the studies have excluded comorbid patients. In the 2nd set of analysis, we allowed comorbidities for cases, i.e. a proportion of patients with disorder A may also have disorder B (and vice versa). We then recorded the summary statistics from the case-control studies with traits A and B against population controls respectively.

For comparison, we also simulated a "real" GWAS comparing the two disorders. More specifically, we considered patients affected with only one disorder but not the other ($N_{A\_only}$ and $N_{B\_only}$). The GWAS comparing $N_{A\_only}$ and $N_{B\_only}$ was regarded as the "real" GWAS in our study. To demonstrate the validity of our current method under sample overlap, we also simulated case-control samples with different overlap rates (*P*). Here, *P* indicates the proportion of overlapped samples among all samples selected for each case-control study, i.e., $P = N_{ctrl.overlap}/2N$. To adjust the overlap rate, we adjust the number of common controls for both traits (as in practice the overlap more often occurs in controls).

**Comparing to genes identified from original GWAS of the two disorders**

Another straightforward approach for finding genes differentially associated with two disorders is to compare the list of significant genes from the original GWAS of trait A vs controls against those from trait B vs controls. To evaluate whether we will uncover the same set of genes or some unique genes may be found by our proposed framework, we carried out analysis on several disorder pairs (MDD vs BPD, SCZ vs BPD,

ADHD vs ASD). We compared the set of differentially associated genes at an FDR cut-off of 0.01 using each of the above approaches.

It should be noted that our proposed statistical framework is different and advantageous in several other aspects, when compared to a 'qualitative' approach of contrasting the genes/variants found in the two original GWAS. Firstly, our approach can provide a formal assessment of the statistical significance or 'confidence' of the differential genes/variants identified. In contrast, to compare the 'significant' genes from the two original GWAS, one needs to set an arbitrary cut-off for the inclusion of genes. Secondly, the proposed method can give an effect size estimate (of the case-control study of trait A vs trait B) of each SNP, which can be further used for downstream analysis like genetic correlations, polygenic scores, transcriptome-wide association studies (TWAS)[25] and so on.

## Results
All supplementary tables are also available at
https://drive.google.com/open?id=1qrpDV6GhobffSwOtRsmAkPY_CihIpHuA

### Simulation results
Table 1 demonstrates our simulation results (please also refer to Table S0). The correlations between the estimated and actual coefficients for the GWAS analysis were very high with different sample sizes of cases. The correlation and RMSE improved with increased sample size and overlap rate (Table 1). Since the sample sizes for current GWAS summary data are usually larger than 10,000, our proposed method should be sufficiently good to approximate the coefficients from GWAS summary data of corresponding traits. As expected, power increases with larger case sizes and heritability explained by SNPs. In addition, there was no observed inflated type I error at a p-value threshold of 0.05.

We also performed additional simulations which allowed subjects to be comorbid for both traits (i.e. a proportion of patients with trait A can also have trait B, and vice versa). The comorbid proportion was set at ~15%. The correlations of the coefficients (beta) and SE were still very high (mostly >0.99) with similar levels of RMSE (Table S0). The type I error was controlled at 5% (at $p<0.05$) while the power was modestly reduced when compared to the simulations under no comorbidities.

### Identification of genetic variants differentiating the psychiatric disorders/traits
For the 18 sets of included GWAS summary statistics (Table 2), we applied the proposed methods to identify differential genetic variants for 26 pairs of comparisons (Table 3). In principle, we selected traits which are similar in nature or commonly comorbid for comparison (please refer to Supplementary Text for details). SNP-based heritabilities are presented in Table 3.

These comparisons may be divided into five groups, including major depressive disorder (MDD) vs. other psychiatric disorders/traits, MDD vs. depression-related traits, neuroticism vs. psychiatric disorders, psychotic experiences vs. three psychiatric disorders and others (Table 3). Altogether, we identify a total of 11,410 significantly associated differential genetic variants ($P<5e-08$) and these variants formed up to 1,398 genomic risk loci based on LD blocks (Table 3).

Here we highlight selected findings, primarily focusing on MDD vs other psychiatric disorders/traits. Please refer to the Supplementary Text for more detailed results and discussions of other comparisons.

**MDD against psychiatric disorders/outcomes**

In this part, we compared MDD with 12 different psychiatric disorders/outcomes, including SCZ, BPD, ED, ASD, ADHD, anxiety disorder, insomnia, alcohol dependence, ever used cannabis, SA, PTSD and OCD (Table 3). Totally 69 genomic risk loci were identified from the 12 pairs of comparisons (Table 3). Please refer to Table S1 to S12 for detailed results.

*MDD against SCZ*

Among the 12 pairs of comparisons, comparison of MDD and SCZ generate the largest number of genome-wide significant SNPs [2,312 SNPs, Table 3 and sub-table 1 in Supplementary Table 1 (Table S1.1)] which belong to 37 genomic risk loci (Table S1.2).

The three gene-mapping strategies (positional, eQTL and CI mapping) generated a set of 524 unique genes, 94 of which were implicated by all three methods (Table S1.5). Additionally, GWGAS analysis identified 953 significant genes (Table 3; Table S1.6). Taken together, 64 genes were implicated by all four strategies. Among them, *CACNA1C* was predicted to have a very high probability of loss-of-function mutation intolerance (pLI score=1; Table S1.5). Genes differentiating MDD and SCZ were mainly enriched in the cortex, the anterior cingulate cortex (BA24), and the frontal cortex (BA9) regions (Table S1.8; FDR<6.0E-04). Cell-type enrichment analysis suggested strong associations with several kinds of neurons in the cortex and prefrontal cortex (Table S1.9). Moreover, this analysis also identified associations with neurons in the midbrain, hippocampus, and lateral geniculate nucleus (LGN) regions (Table S1.9). Conditional analyses suggested neurons in the cortex, GABAergic neurons in the midbrain, and pyramidal neurons in the hippocampus as *independent* contributing neurons (after controlling for other cell types) (Table S1.10).

In gene-set enrichment analysis (GEA), the 953 GWGAS significant genes were enriched in a number of biological GO sets, including generation of neurons, regulation of nervous system development and central nervous system neuron differentiation (Table S1.11). We also conducted genetic correlation analysis in which SCZ was defined as the 'case' and MDD as (pseudo-)'controls'. Note that a positive genetic correlation indicates that the 'case' disorder is more positively associated with the studied trait genetically than the (pseudo-)'control' disorder, and vice versa. For example, we observed inverse genetic correlations (rg) with insomnia, neuroticism, coronary artery disease (CAD) and mean hippocampal volume, among others. This suggested that MDD has stronger positive genetic correlations with the above traits/disorders compared to SCZ. Findings of this type may shed light on different patterns of comorbidities, but may also be clinically informative. For instance, the significant inverse rg with CAD suggested that compared to SCZ patients, MDD patients may be more genetically predisposed to CAD.

*MDD against BPD, ED, ASD, ADHD, Anxiety disorder, Insomnia, Alcohol dependence and Cannabis use*

In these 8 pairs of comparisons, we identified 32 differential genomic loci (Table 3 and S2-S9). The comparison between MDD and BPD revealed the largest number of significant genes based on GWGAS (174 genes; Table 3; Table S2.6). Details are presented in Tables S2-S9 and the Supplementary Text.

**MDD against depression-related traits**

Here we tried to identify differential genetic variants from three pairs of comparisons between MDD and three depression-related phenotypes (probable recurrent severe depression, seen GP for anxiety/depression and longest period of low/depressed). Ten risk loci were identified (Table 3/Tables S13-S15).

*MDD against depression defined in UKBB*

First we compared MDD (from PGC; majority clinically defined) against probable recurrent major depression (severe) (ProbDep). We identified 4 risk loci (Table 3; Table S13.2), including one in the extended MHC(xMHC) region. Gene-based test revealed 110 significant genes. Tissue enrichment analysis highlighted the cerebellar hemisphere, nucleus accumbens and frontal cortex as the most enriched regions. Cell-type enrichment analysis suggested that the significant genes were associated with GABAergic, dopaminergic and other types of neurons in LGN, middle temporal gyrus (MTG), hippocampus, midbrain and cortex regions (Table S13.9). Genetic correlation analysis showed that MDD-PGC was more positively genetically correlated with most other psychiatric disorders (e.g. SCZ/BPD/ASD/ADHD) as well as CAD when compared with ProbDep (Table S13.13). We also compared MDD against 'seen GP for nerves/anxiety/depression' (GPDep), with detailed results shown in Table S14.

*MDD against duration of longest period of feeling low/depressed (top quintile as case)*

Three genetic risk loci were identified, including the *GRIK2* gene. The gene codes the Glutamate Ionotropic Receptor Kainate Type Subunit 2, suggesting glutamatergic transmission may be one factor with differential associations between susceptibility to depression and severity (as reflected by duration) of illness.

**Neuroticism against SCZ/MDD/Anxiety disorder/alcohol dependence**

These comparisons were made based on relatively high association of neuroticism with these disorders[26-28]. We identified 1,294 genomic risk loci (Table 3). Please refer to Tables S16-S19 for details.

**Psychotic experiences against SCZ/BPD/MDD**

We identified 10 and 2 genomic risk loci from comparison of psychotic experiences against SCZ and BPD respectively, but not from psychotic experiences against MDD (Table 3/Tables S20-S22).

**Other comparisons**

We also applied the proposed methods to SCZ against BPD, ADHD against ASD, alcohol dependence against ever used cannabis and anxiety disorder against SA (Table 3). We identified 3, 7, 2 and 1 genomic risk loci from each of the comparison respectively. Please refer to the Supplementary Text and Tables S23-S26 for details.

**Distinguishing between disorders based on PRS**

*Potential ability to distinguish between disorders based on PRS derived from current GWAS data*

In the analysis, we assume each subject is either having one of the disorders. Taking SCZ vs MDD as an example, we assume the differential diagnosis (ddx) has been narrowed down to either SCZ or MDD. The prior probabilities (without genetic information) of being affected with either disorder are based on lifetime prevalence of the disorders[29, 30]. For example, here we assume a person has ~13/0.5=26 times of being affected by MDD than SCZ, in the absence of additional information. Our analytic framework actually allows more flexible setting of these prior probabilities, although we made simpler assumptions here. We expect that with the addition of polygenic scores, one would be able to differentiate the disorders more accurately. A good prediction model leads to more spread-out predicted risks and larger relative risks when we compare subjects at the top and bottom percentiles.

Subjects at the lowest $5^{th}$ percentile of the PRS distribution (SCZ as 'case' and MDD as 'control') have markedly lower risks of SCZ than MDD compared to the population average. In this case, relative risk (RR) of MDD vs SCZ was 125.3 for a person with PRS at the bottom $5^{th}$ percentile (average RR=26) (Table S1.15). With an increase in PRS, the risk of SCZ became higher, while the risk of MDD reduced. Subjects at the highest 5% of the risk score (of SCZ vs MDD) had a substantially decreased RR of 10.16. At the start we assume ~26 times higher risks of MDD than SCZ based on overall lifetime risks; a reduction to 10.16 times is a relatively large change. We also present the RR of the 'case' disorder by comparing individuals at the highest and lowest $x$th percentiles. For example, the estimated RR of SCZ was 11.31 if we compare those at the highest $5^{th}$ against those at the lowest $5^{th}$ percentile. For SCZ vs BPD and BPD vs MDD, the corresponding RR (for the $1^{st}$ disorder) was 3.29 and 2.82 respectively.

For most comparisons, the AUC based on PRS of *existing* GWAS data were modest, with several pairs showing AUC>0.6. For example, we estimated that AUCs for distinguishing SCZ from MDD, BPD from MDD and SCZ from BPD were 0.694, 0.602 and 0.618 respectively. The AUCs of distinguishing SCZ *vs* psychotic experience and ADHD *vs* ASD were estimated to be 0.686 and 0.622 respectively.

*Maximum AUC based on SNP-based heritability (i.e. all GWAS-panel variants)*
The maximum AUC attainable (at SNP-based heritability) is presented in Table 5. The levels were much higher than the current AUC, indicating room for improving discriminating ability by increasing sample sizes. For example, based on SNP-based heritability, the AUCs for distinguishing SCZ from MDD, BPD from MDD and SCZ from BPD were 0.763, 0.749 and 0.726 respectively. We also computed other predictive indices/graphs which are shown in supplementary tables.

**Comparing to genes identified from original GWAS of the two disorders**
The results are presented in Table S27-29. Briefly, in the comparison of ASD vs ADHD, 40 genes had FDR<0.01 based on our analysis, of which 27 did not overlap with the genes found by simple comparison of the original GWAS (at FDR<0.01). For BPD vs MDD, the corresponding numbers were 38 and 5; for SCZ vs BPD, the corresponding numbers were 49 and 15.
We note that the two approaches are different and the results are not directly comparable, as directly comparing genes found from the two original GWAS does not provide a formal assessment of the statistical significance of individual genes. The results are shown for reference and as an exploratory analysis.

**Discussion**

The present study applied a simple yet useful analytic framework to identify differential genetic markers for a board range of psychiatric disorders/traits. We conducted detailed secondary analysis to identify the genes, pathways and cell-types/tissues implicated. From the 26 pairs of comparisons, we identified a total of 11,410 significantly associated differential variants, 1,398 genomic risk loci and 3,362 significant genes from GWGAS with FDR<0.05.

*SNP-based heritability($h^2_{snp}$) of differential genetic associations*

Here found that the SNP-based heritabilities were significantly different from zero for almost all comparisons between psychiatric disorders, with some having moderately high heritabilities. This suggests that genetic differences (due to common variants) may at least partially underlie the differences in susceptibility between psychiatric disorders, even for closely related ones such as MDD and anxiety disorders.

For MDD and comparisons with other disorders, we observed the highest $h^2_{snp}$ in the comparison with SCZ, BPD, ED and anxiety disorders. For instance, despite substantial symptom overlap[31] between MDD and anxiety disorders, the $h^2_{snp}$ is among the highest at ~36% (by SumHer; on liability scale). On the other hand, $h^2_{snp}$ was estimated at ~1% only when comparing MDD to PTSD. A possible explanation is that environmental factors (e.g. traumatic stressors must be present for PTSD but not MDD) may play an important role in explaining the differences between the two disorders. For neuroticism against other psychiatric disorders, the $h^2_{snp}$ were in general low; however, $h^2_{snp}$ for neuroticism itself was only ~10%[32].

We wish to highlight a difference between genetic correlation (between two traits) and the $h^2_{snp}$ from the differential association test. Two variables can have a high correlation if there is a strong linear relationship, but the actual values of the variables can differ. It is possible that two traits have a high genetic correlation (rg), but as the effect sizes of SNPs can differ, $h^2_{snp}$ can still be substantial. There are several caveats when interpreting $h^2_{snp}$. First, large samples are often required for $h^2_{snp}$ analysis. However, for several disorders sample sizes were relatively moderate (e.g. OCD/ED); as such the estimates could be imprecise. Also, contribution of rare variants and other 'omic' changes (e.g. epigenetic changes) were not captured by $h^2_{snp}$. Moreover, estimation of $h^2_{snp}$ is subject to model assumptions[20] of genetic architecture, which can vary across diseases.

*Potential ability of PRS from existing GWAS data to differentiate disorders*

A potential translational aspect is to make use of PRS from SNPs to distinguish between psychiatric disorders, which has been raised, for example, by ref[33] in a recent review. This is particularly relevant in psychiatry due to the lack of objective biomarkers. In an earlier work, Hamshere et al.[34] found that PRS of SCZ was able to differentiate schizoaffective BPD patients from non-schizoaffective BPD patients. In a more recent study, Liebers et al.[35] studied if PRS may discriminate BPD from MDD. They found that subjects at the top decile of BPD PRS were significantly more likely to have BPD than MDD, when compared to those in the lowest decile. The estimated odds ratio (OR) was 3.39 (95% CI 2.19–5.25), which is comparable to our relative risk estimate (2.24) (see Table 5; RR are usually smaller than ORs).

Among the comparisons, DDx between MDD and BPD is one of the most clinically relevant. Based on present GWAS data, the AUC for discriminating BPD vs MDD is 0.602 (at the best p-value threshold), which is modest. However, PRS may be more informative for individuals at the extreme end of the score. The discriminating power between BPD and SCZ was similarly modest (best AUC=0.618) but the AUC for SCZ vs MDD was much higher (0.694). Clinically, major depression (mainly psychotic depression) may be a DDx for first-episode psychosis[36]; it may be interesting to study if PRS can help distinguish SCZ from MDD in such patients. We also estimated the maximum discriminatory ability by PRS based on $h^2_{snp}$ (i.e. assuming all common variants are found); the maximum AUC for MDD vs BPD, SCZ vs BPD and SCZ vs MDD were 0.749, 0.726 and 0.763 respectively. These findings suggest that with larger GWAS sample sizes, PRS may become more informative and may help DDx. Another interesting analysis is on how well PRS can differentiate 'psychotic experience'(PE) against psychiatric disorders such as SCZ, BPD and MDD, which we found high AUC based on $h^2_{snp}$ but poorer discriminatory power using existing GWAS data (see Supplementary Text).

Several limitations are worth noting. For some comparisons (e.g. MDD vs other disorders), comorbidities are possible. For our PRS-based DDx, as stated before, it was assumed that the subject is either having one of the disorders. For example, we assume the DDx has been narrowed down to either SCZ or MDD in our analysis of SCZ vs MDD. In practice, the above assumption may be true for some disorder pairs (e.g. SCZ vs BPD; BPD vs unipolar depression) but may not hold for others (for which a patient can have both disorders at the same time). As such, the PRS analysis results and AUCs should be viewed with caution, although we believe they are still of scientific interest. On the other hand, in the presence of comorbidities, the PRS approach may still be able to inform whether a person has a higher genetic predisposition to one disorder than the other, say SCZ compared to MDD. Whether this may be of importance clinically will require further studies. For example, an interesting question is that whether a specific treatment (e.g. for disorder *A*) may be more effective in patients with higher genetic predisposition to disorder *A* (or vice versa), even if comorbidity is possible.

In practice, we expect clinical symptoms and features still remain very important in making DDx. Genomic data may provide additional discriminatory power when integrated with clinical features. Also, since we relied on summary statistics, we applied an analytic approach[24] to estimate the AUC from current GWAS samples. Limitations of this methodology were detailed in[24]. Mainly, we assume the predictive model will be applied to the same population as the training data. Nevertheless, as patients with the same psychiatric disorder can be heterogeneous, and PRS may need to be applied across different ethnic groups, the estimated AUCS may be optimistic in this regard. Ideally, predictive power should be further evaluated in an independent set with individual genotype data. In addition, our analytic approach for forecasting AUC assumed a (standard) p-value thresholding and LD-clumping(P+T) approach. This approach is widely adopted, but newer PRS modelling methodologies (e.g. LDpred; see [37] for a review) may be used to further improve predictive power.

*Comparison of MDD-PGC with depression-related traits in UKBB*
We performed another interesting comparison between MDD-PGC[9] and other depression-related traits from UKBB. The former group was mainly composed of clinically diagnosed MDD, while the latter group was largely defined by self-reporting. For example, for recurrent probable major depression (severe) (ProbDep), it included subjects who reported feeling depressed for one week, with >=2 episodes for >=2 weeks, and have visited a psychiatrist (please also refer to Smith et al[10] and https://biobank.ctsu.ox.ac.uk/crystal/crystal/docs/MentalStatesDerivation.pdf). The other phenotype was having seen GP for depression/nerves/anxiety. Neither trait involved assessment of clinical symptoms as described in DSM/ICD. Based on our analysis, MDD-PGC appeared to be more strongly genetically correlated with other psychiatric disorders (e.g. SCZ/BPD/anorexia/ASD/ADHD) and other outcomes such as CAD, when compared with non-clinically-defined depression in UKBB. Interestingly, while rg between MDD-PGC and UKBB depression traits were high, the SNP-based heritability from differential genetic analysis was significantly larger than zero. One possible explanation is that while many susceptibility genes may be shared between them, the effect sizes may differ. Another point to note is that rg based on LDSC may be over-estimated in case-control studies due to difficulties in handling covariates[38]. The latter has been reported in[39] when comparing LDSC against a more sophisticated method PCGC[38].

Recently Cai et al.[39] suggested that depression traits defined by 'minimal phenotyping' (ProbDep and GPDep included here also belonged to 'minimal phenotyping') are genetically different from strictly defined MDD. For example, they have lower $h^2_{snp}$ and have worse predictive power in MDD cohorts. Cai et al. focused on comparisons of different definitions of 'depression' within UKBB, while here we mainly compared the genetic architecture of MDD-PGC against traits in UKBB; we also employed a different statistical approach. Our results supported differences in genetic basis between different definitions of depression, and calls for more in-depth phenotyping to study depression and related traits.

We would like to highlight an important limitation of the above comparisons. Although the MDD-PGC sample mainly comprises clinically-diagnosed MDD, several limitations cannot be addressed in this analysis. For example, the proportion of patients with comorbid disorders is unknown. Also, the diagnostic approach, inclusion/exclusion criteria, clinical features and other sample characteristics may differ across sub-studies.

*Tissue/cell-type enrichment analysis*
In view of the large number of comparisons performed, we just highlighted a few results for discussion. The tissue/cell-type enrichment analysis implied that the frontal cortex (BA9) and anterior cingulate cortex (ACC; BA24) may be implicated in the *difference* between several disorders, such as MDD against SCZ/BPD, neuroticism against MDD/alcohol dependence, and psychotic experiences against SCZ/BPD. BA9 contributes to the dorsolateral and medial prefrontal cortex, dysfunction of which underlies many cognitive and behavioural disturbances that are associated with psychiatric disorders, such as SCZ, MDD, ADHD, and ASD [40, 41]. The ACC is involved in many functional roles of the brain, including affective, cognitive and motor aspects [42]. A number of studies have suggested that functioning alterations in the ACC may be implicated in psychiatric disorders such as MDD[43], BPD[44] and SCZ[45]. It is possible that different patterns of dysfunctioning in these brain regions may underlie the differences between the disorders.

Cell-type enrichment analysis may also help to pinpoint the cell-types (and brain regions) involved in differentiating the disorders. For example, when comparing MDD vs anxiety disorders, the most enriched cell types were GABAergic neurons from hippocampus, midbrain and temporal cortex. Interestingly, benzodiazepines, one of the most widely prescribed drugs for anxiety, acts on the GABAergic pathway, while antidepressants primarily target the monoamine system. With increasing amount of single-cell RNA-seq data in the future, cell-type enrichment analysis may delineate more precisely the specific type of neurons involved.

*Genetic correlation analysis*

We just briefly highlight a few examples of our findings. For example, in the comparison of BPD vs MDD, we observed positive genetic correlation(rg) with childhood intelligence and level of education (Table S2). This suggests that BPD, when compared to MDD, is more strongly genetically linked to these traits (in the positive direction). This is in line with a previous study which reported that low intelligence was associated with severe depression and SCZ but not BPD [46]. On the other hand, another study reported that in men with no psychiatric comorbidity, both low and high intelligence are risk factors for BPD[47]. However, these are epidemiological studies and further studies are required to validate our findings and to reconcile with epidemiological findings. Slightly unexpectedly, we also observed positive rg of BPD vs MDD with anorexia nervosa (AN), although AN is more commonly comorbid with depression clinically. Similarly, when comparing SCZ vs MDD, positive rg with AN was also observed. This could suggest that MDD in AN is more strongly influenced by environmental factors [48].

Genetic correlation analysis may also shed light on the brain regions implicated. For the comparison of SCZ vs MDD, we observed a significant negative correlation with hippocampal volume, which was corroborated by associations in the hippocampus region in cell-type enrichment analysis. Both SCZ and MDD patients were reported to have smaller hippocampal volumes compared to healthy controls[49, 50]. However, a comparative study showed that there was a larger reduction in hippocampal volumes in SCZ compared to MDD[51]. Our rg analysis appeared to support this finding.

*Other limitations*

Many limitations of this study have been detailed above. As for other limitations, we note that the methodology assumes the controls of both GWAS datasets originate from a similar population. If the heterogeneity is high (e.g. from different ethnic groups), the estimates may be biased. As a related limitation, most of the studies were based on Europeans; the effects of genetic loci may differ across populations, and PRS derived from Europeans may have poorer predictive abilities in other ethnicities.

Besides, information on possible comorbidities is not available from most GWAS datasets. For example, in a comparison of MDD vs OCD, some OCD patients may have comorbid MDD. To a certain extent, this is similar to the use of unscreened controls in genetic studies, which may lead to reduction of power to detect genetic variants but generally does not increase risks of false positives[52, 53]. Our simulations also supported the validity of the proposed statistical approach in the presence of comorbidities. However, from a clinical perspective, comorbidities also affect the interpretation of PRS-based differentiation of disorders, which has been discussed in detail above.

*Conclusions*

Our analytic framework successfully identified a number of differential genomic risk loci from 26 pairs of comparisons of psychiatric traits/disorders. Moreover, further analysis revealed many novel genes, pathways, brain regions and specific cell types implicated in the differences between disorders. We also showed that PRS may help differentiation of some psychiatric disorders to a certain extent, but further clinical studies are required.

**Supplementary materials**

**Please note that all supplementary tables are available at**

https://drive.google.com/open?id=1qrpDV6GhobffSwOtRsmAkPY_CihIpHuA

**Author contributions**

Conceived and designed the study: HCS. Supervised the study: HCS. Methodology: HCS (lead), LY. Data analysis: SR (lead), LY, with input from YX. Simulation experiments: LY. Data interpretation: SR, LY, HCS. Writing of manuscript: HCS, SR, with input from LY.

**Acknowledgements**

We would like to thank Prof. Stephen Tsui for computing support. This study was partially supported by a National Natural Science Foundation of China (NSFC) grant for Young Scientist (31900495), the Lo Kwee Seong Biomedical Research Fund and a Chinese University of Hong Kong Direct Grant. We thank Mr. Carlos Chau for assistance in part of the analyses, and Ms Jinghong Qiu for editing the manuscript.

**Conflicts of interest**

The authors declare no relevant conflicts of interest.

Figure 1 legends

Summary of our analysis framework. GWAS summary statistics of the two traits under study were harmonized and differential genetic associations were identified by the method we described in main text. The power of polygenic scores (derived from the above association test treating the first disorder as 'case' and the second one as 'control') to differentiate the two disorders was computed. We computed two sets of discriminatory power estimates, one based on *existing* GWAS data, the other based on SNP-based heritability, reflecting the maximum achievable discriminatory power. We also investigated the genetic correlation (mainly using LD score regression [LDSC]) with other possible comorbid traits/disorders. We also performed genome-wide gene-based association study (GWGAS) to identify associated genes, and the most relevant tissues, cell-types and pathways implicated. As a parallel analysis, we performed functional annotations and mapped the SNPs to relevant genes based on gene positions, expression quantitative trait loci (eQTL) and chromatin interaction (CI) data.

**Tables**

Table 1 Simulation results comparing analyses of individual-level genotype data and our presented analytic approach

| Overlap rate | No. cases | $h^2_A$ | $h^2_B$ | Correlation | | RMSE | | Inferred | | Real GWAS | |
|---|---|---|---|---|---|---|---|---|---|---|---|
| | | | | Beta | SE | Beta | SE | Power | Type I error | Power | Type I error |
| 0.15 | 10000 | 0.2 | 0.3 | 0.98769 | 0.99789 | 0.02194 | 0.00803 | 0.633 | 0.040 | 0.723 | 0.043 |
| | 20000 | 0.2 | 0.3 | 0.99335 | 0.99807 | 0.01634 | 0.00564 | 0.740 | 0.040 | 0.770 | 0.037 |
| | 50000 | 0.2 | 0.3 | 0.99766 | 0.99811 | 0.00939 | 0.00359 | 0.823 | 0.023 | 0.873 | 0.047 |
| | 100000 | 0.2 | 0.3 | 0.99861 | 0.99811 | 0.00724 | 0.00253 | 0.877 | 0.047 | 0.903 | 0.047 |
| | 10000 | 0.22 | 0.32 | 0.98766 | 0.99776 | 0.02253 | 0.00800 | 0.653 | ------ | 0.723 | ------ |
| | 20000 | 0.22 | 0.32 | 0.99376 | 0.99794 | 0.01618 | 0.00565 | 0.723 | ------ | 0.787 | ------ |
| | 50000 | 0.22 | 0.32 | 0.99784 | 0.99792 | 0.00941 | 0.00360 | 0.833 | ------ | 0.880 | ------ |
| | 100000 | 0.22 | 0.32 | 0.99873 | 0.99796 | 0.00724 | 0.00254 | 0.877 | ------ | 0.910 | ------ |
| 0.25 | 10000 | 0.2 | 0.3 | 0.99022 | 0.99768 | 0.02077 | 0.00606 | 0.660 | 0.040 | 0.717 | 0.043 |
| | 20000 | 0.2 | 0.3 | 0.99645 | 0.99764 | 0.01243 | 0.00437 | 0.737 | 0.037 | 0.770 | 0.030 |
| | 50000 | 0.2 | 0.3 | 0.99816 | 0.99785 | 0.00898 | 0.00275 | 0.817 | 0.043 | 0.857 | 0.050 |
| | 100000 | 0.2 | 0.3 | 0.99913 | 0.99777 | 0.00608 | 0.00194 | 0.870 | 0.040 | 0.900 | 0.027 |
| | 10000 | 0.22 | 0.32 | 0.99144 | 0.99730 | 0.02031 | 0.00605 | 0.683 | ------ | 0.727 | ------ |
| | 20000 | 0.22 | 0.32 | 0.99637 | 0.99754 | 0.01315 | 0.00436 | 0.757 | ------ | 0.780 | ------ |
| | 50000 | 0.22 | 0.32 | 0.99831 | 0.99771 | 0.00888 | 0.00275 | 0.833 | ------ | 0.860 | ------ |
| | 100000 | 0.22 | 0.32 | 0.99923 | 0.99760 | 0.00597 | 0.00194 | 0.887 | ------ | 0.903 | ------ |

No. cases indicates the number of cases we defined for our simulation scenarios; $h^2$, heritability explained by SNPs; RMSE, root mean square error.

**Table 2. Summary of GWAS data of 18 psychiatric traits/disorders included in this study**

| Traits/Disorders | Abbreviation | Source [b] | Data type | Cases | Controls | Total N | Prevalence (%) [d] |
|---|---|---|---|---|---|---|---|
| Major Depression Disorder (2018)[9] | MDD | PGC | binary | 59,851 | 113,154 | 173,005 | 13.0 [54] |
| Bipolar Disorder (2018)[8] | BPD | PGC | binary | 20,129 | 21,524 | 41,653 | 2.4 [55] |
| Schizophrenia (2018)[17] | SCZ | Pardiñas et al. [17] | binary | 40,675 | 64,643 | 105,318 | 0.5 [29] |
| Autism Spectrum Disorder (2019)[56] | ASD | iPSYCH&PGC | binary | 18,381 | 27,969 | 46,350 | 2.5 [57] |
| Attention deficit hyperactivity disorder (2019)[58] | ADHD | iPSYCH&PGC | binary | 19,099 | 34,194 | 53,293 | 6.5 [59] |
| Post-traumatic Stress Disorder (2017)[52] | PTSD | PGC | binary | 5,183 | 15,547 | 20,730 | 3.9 [60] |
| Anxiety Disorder (2018) [a] | anxiety | UKBB | binary | 16,730 | 101,021 | 117,751 | 14.2 [e] |
| Eating Disorder (2017)[61] | ED | PGC | binary | 3,495 | 10,892 | 14,477 | 1.2 [62] |
| Obsessive-compulsive Disorder (2018)[63] | OCD | PGC | binary | 2,688 | 7,037 | 9,725 | 2.3 [64] |
| Insomnia (2019)[65] | Insomnia | UKBB&CTG lab | binary | 109,389 | 277,144 | 386,533 | 10.0 [66] |
| Suicide Attempts in mental disorder (2018) [67] | SA | iPSYCH-PGC | binary | 6,024 | 44,240 | 50,264 | 2.7 [68] |
| Alcohol dependence (2018)[69] | alcohol | PGC | binary | 11,476 | 23,080 | 34,556 | 12.0 [70] |
| Ever used cannabis (2018)[71] | cannabis | ICC | binary | 43,380 | 118,702 | 162,082 | 4.0 [72] |
| Psychotic Experiences (2019) [73] | PE | CNGG-Walters group | binary | 6,123 | 121,843 | 127,966 | 5.8 [74] |
| Neuroticism_Highest_20% (2019) [75] | neuroticism | CTG Lab | binary [c] | 78,056 | 312,222 | 390,278 | 20.0 [c] |
| Longest period of depression_Highest_20% | longest depression | UKBB | binary [c] | 10,133 | 40,531 | 50,664 | 20.0 [c] |
| Probable Recurrent major depression (severe) | ProbDep | UKBB | binary | 6,304 | 80,591 | 86,895 | 3.55 [76] |
| Seen doctor (GP) for nerves, anxiety, tension or depression | GPDep | UKBB | binary | 123,528 | 235,165 | 358,693 | 17.3 [54] |

[a] Anxiety Disorder: mental health problems ever diagnosed by a professional: anxiety, nerves, or generalized anxiety disorder; [b] PGC: psychiatric genomics consortium; UKBB: UK biobank; CTG lab: complex trait genetics lab; ICC: international cannabis consortium; CNGG: Centre for Neuropsychiatric Genetics and Genomics; [c] continuous neuroticism scores were transformed to binary traits, in which 20% of subjects were assigned as cases and the others as controls; population prevalence of longest depression is population prevalence of MDD multiple by 20%, i.e. 13%*20%. [d] Prevalence of traits/disorders refers to estimates of lifetime prevalence based on the cited references; [e] Prevalence of anxiety disorder is estimated from UKBB directly.

**Table 3. Identification of differentially associated genetic variants/genes from correlated psychiatric traits/disorders.**

| Comparisons [a] | Genetic correlation | | | Differential association GWAS[b] | | | | |
|---|---|---|---|---|---|---|---|---|
| | $rg$ | p-value | intercept | Sig. SNPs | Genomic risk loci | Sig. Genes | LDSC-$h^2$ (se) | SumHer-$h^2$ (se) |
| 1. MDD vs. psychiatric disorders/traits | | | | | | | | |
| SCZ | 0.3857 | 3.50e-46 | 0.0548 | 2,312 | 37 | 953 | 0.183(0.008) | 0.220 (0.085) |
| BPD | 0.3387 | 1.82e-23 | 0.0679 | 42 | 4 | 174 | 0.239(0.013) | 0.297 (0.040) |
| ED | 0.1652 | 1.29e-02 | 0.0433 | 94 | 2 | 5 | 0.258(0.035) | 0.319 (0.067) |
| ASD | 0.4466 | 6.97e-25 | 0.1441 | 76 | 5 | 17 | 0.147(0.015) | -[d] |
| ADHD | 0.5573 | 1.33e-50 | 0.1703 | 167 | 5 | 82 | 0.198(0.018) | 0.162 (0.090) |
| anxiety | 0.7851 | 1.87e-32 | 0.0341 | 120 | 5 | 106 | 0.284(0.019) | 0.363 (0.035) |
| insomnia | 0.4706 | 1.75e-44 | 0.004 | 13 | 3 | 40 | 0.083(0.006) | 0.099 (0.011) |
| alcohol | 0.5893 | 4.23e-09 | 0.0389 | 3 | 1 | 0 | 0.064(0.016) | 0.088 (0.058) |
| cannabis | 0.2433 | 8.61e-09 | -0.0009 | 608 | 7 | 24 | 0.122(0.009) | 0.133 (0.016) |
| SA | 0.5639 | 8.00e-04 | 0.0069 | 0 | 14[c] | 0 | 0.056(0.023) | 0.117 (0.048) |
| PTSD | 0.6095 | 7.70e-03 | 0.0064 | 0 | 9[c] | 0 | 0.034(0.024) | 0.010 (0.048) |
| OCD | 0.2272 | 5.00e-04 | 0.0103 | 0 | 12[c] | 1 | 0.344(0.050) | 0.262 (0.081) |
| 2. MDD vs. depression-related traits (from UKBB) | | | | | | | | |
| Recurrent probable depression | 1.1036 | 7.99e-12 | 0.0617 | 177 | 4 | 110 | 0.505(0.033) | 0.613 (0.055) |
| seen GP for depression | 0.9441 | 6.84e-287 | 0.0978 | 22 | 3 | 72 | 0.087(0.007) | 0.111 (0.012) |
| longest depression | 1.0821 | 1.21e-06 | 0.0286 | 3 | 3 | 0 | 0.020(0.003) | 0.021 (0.006) |
| 3. Neuroticism vs. psychiatric disorders | | | | | | | | |
| anxiety | 0.7401 | 7.24e-49 | 0.1617 | 4,671 | 57 | 1,232 | 0.161(0.006) | 0.259 (0.076) |
| SCZ | 0.2293 | 1.10e-20 | 0.0102 | 1,733 | 1,214 | 1 | 0.013(0.001) | 0.009 (0.003) |
| MDD | 0.7507 | 6.54e-182 | 0.0644 | 65 | 20 | 4 | 0.016(0.001) | 0.021 (0.001) |
| alcohol | | 0.3754 | 7.66e-08 | 0.008 | 37 | 3 | 40 | 0.022(0.004) | 0.105 (0.004) |
| 4. Psychotic experiences vs. common disorders | | | | | | | | | |
| SCZ | | 0.2014 | 1.00e-04 | 0.009 | 731 | 10 | 239 | 0.356(0.022) | 0.330 (0.043) |
| BPD | | 0.1748 | 2.36e-02 | 0.0056 | 25 | 2 | 30 | 0.398(0.033) | 0.587 (0.083) |
| MDD | | 0.4957 | 1.33e-08 | 0.0072 | 0 | 11 [c] | 0 | 0.131(0.022) | - [d] |
| 5. Four other comparisons | | | | | | | | | |
| SCZ vs. BPD | | 0.6903 | 2.13e-181 | 0.1403 | 45 | 3 | 144 | 0.202(0.015) | 0.378 (0.071) |
| ASD vs. ADHD | | 0.3879 | 4.89e-17 | 0.3444 | 335 | 7 | 88 | 0.400(0.033) | 0.438 (0.063) |
| alcohol vs. cannabis | | 0.1482 | 5.20e-02 | 0.022 | 130 | 2 | 0 | 0.132(0.019) | 0.187 (0.027) |
| anxiety vs. SA | | 0.0705 | 7.31e-01 | 0.0068 | 1 | 1 | 0 | 0.072(0.036) | 0.137 (0.065) |

[a] MDD: major depression disorder, SCZ: schizophrenia, BPD: bipolar disorder, ED: eating disorder (anorexia nervosa), PTSD: posttraumatic stress disorder, OCD: obsessive-compulsive disorder, ASD: autism spectrum disorder, ADHD: attention deficit hyperactivity disorder; Used cannabis: ever used cannabis; Longest depression: Longest period of depression; Seen GP for depression: Seen general practitioner (GP) for nerves, anxiety, tension or depression; Severe depression: Bipolar and major depression status: Probable Recurrent major depression; SA: suicide attempts.

[b] Sig. SNPs: SNPs with nominal p values below 5e-08; Sig. Genes: Genes with adjusted p-value (FDR) below 0.05 in GWGAS; LDSC/SumHer-$h^2$: liability-scale SNP-based heritability calculating by the LDSC and SumHer programs, respectively.

[c] The corresponding genomic risk loci are constructed on those lead SNPs with *p-value below 5e-06*. Values of AUC above 0.70 are in bold.

[d] SumHer returns estimates that are negative, hence we present the results from LDSC only.

Table 4. Top 5 genes from each comparison based on gene-based analysis using MAGMA.

| Comparison | Top 5 genes | $P$ | FDR-adjusted $P$ |
|---|---|---|---|
| SCZ vs MDD | **PPP1R16B, HIST1H4L, DPYD, PITPNM2, NGEF** | <5.44E-12 | <1.99E-08 |
| BPD vs MDD | **HAPLN4, TRANK1, VPS9D1, MAD1L1, NDUFA13** | <4.24E-07 | <1.33E-03 |
| MDD vs ASD | **MACROD2, XRN2, WDPCP, EGR2, FZD5** | <4.00E-06 | <1.27E-02 |
| MDD vs ADHD | **CDH8, MEF2C, KDM4A, PTPRF, KCNH3** | <2.02E-07 | <7.48E-04 |
| MDD vs ED | **ERBB3, SUOX, FAM19A2, CRTC3, RAB5B** | <5.58E-06 | <2.09E-02 |
| MDD vs anxiety | **BTN3A2, HIST1H2BN, PTPN1, ZKSCAN4, PGBD1** | <3.54E-08 | <1.33E-04 |
| MDD vs. insomnia | **BTN3A2, HIST1H2BN, ZSCAN9, SYNGAP1, RAB1B** | <4.98E-07 | <3.81E-03 |
| MDD vs. alcohol | MTFR1, ATF6B, KREMEN2, SLC25A52, ALPK1 | <1.35E-04 | <4.47E-01 |
| MDD vs. cannabis | **CADM2, C10orf32-ASMT, AS3MT, ACTL8, ARID1B** | <3.80E-07 | <1.43E-03 |
| MDD vs. SA | NCL, ST8SIA5, COA4, PDE4B, SLBP | <1.46E-04 | <3.99E-01 |
| MDD vs PTSD | ATP6V1E1, MYO5B, ZYG11A, GNA15, UBA3 | <3.04E-04 | <8.89E-01 |
| MDD vs OCD | **KIT**, PLAG1, FGF19, PPIG, TXNL1 | <3.31E-05 | <8.88E-02 |
| MDD vs. severe depression | **HIST1H2BN, BTN3A2, ZKSCAN4, PGBD1, PTPN1** | <3.89E-08 | <1.47E-04 |
| MDD vs. seen GP for depression | **PTPN1, BTN3A2, ZKSCAN4, HIST1H2BN, PGBD1** | <1.54E-07 | <5.81E-04 |
| Longest depression vs. MDD | FBXW4, C11orf42, AC079602.1, ANAPC11, HIST1H2BM | <1.56E-03 | <8.33E-01 |
| Neuroticism vs. anxiety | **STH, WNT3, SPPL2C, CRHR1, MAPT** | <4.16E-22 | <1.57E-18 |
| Neuroticism vs. SCZ | **DNAJC19**, CCL20, OR7D4, PBX2, REG3A | <5.65E-05 | <2.01E-01 |
| Neuroticism vs. MDD | **MAPT, WNT3, CRHR1, KANSL1**, NSF | <1.73E-05 | <6.50E-02 |
| Neuroticism vs. alcohol | **BTN3A2, HIST1H2BN, ZSCAN9, SYNGAP1, RAB1B** | <4.98E-07 | <3.81E-03 |
| Psychotic experiences vs. SCZ | **SPATS2L, HIST1H4L, UBD, OR2B2, HIST1H2BN** | <6.47E-10 | <2.02E-06 |
| Psychotic experiences vs. BPD | **SPATS2L, HAPLN4, TM6SF2, NDUFA13, CTC-260F20.3** | <3.73E-07 | <1.24E-03 |
| Psychotic experiences vs. MDD | FAM168A, SHPRH, SPAM1, ADRB2, POC1B | <1.48E-04 | <3.66E-01 |
| SCZ vs BPD | **ZKSCAN3, ZSCAN31, PGBD1, HYDIN, ZSCAN12** | <6.52E-08 | <2.38E-04 |
| ADHD vs ASD | **XKR6, KDM4A, C8orf12, RP1L1, MSRA** | <1.52E-08 | <5.65E-05 |
| Cannabis vs alcohol | HS6ST1, ABHD14A-ACY1, ACY1, ENTPD4, AKR1C2 | <6.81E-05 | <2.10E-01 |
| Anxiety vs SA | PDE4B, DNAJC6, LAMA2, NCL, RAVER1 | <3.96E-05 | <1.45E-01 |

Genes with FDR<0.05 are in bold.

Table 5. Ability to discriminate psychiatric disorders by polygenic risk scores (PRS)

| Comparison | Discriminating Ability | | | Relative risk of the 1st disorder comparing the top against the bottom percentiles (based on existing GWAS) | | | |
|---|---|---|---|---|---|---|---|
| | AUC (Max) | Polygenic risk scores (based on existing GWAS) | | Top 5th vs. lowest 5th percentile | Top 10th vs. lowest 10th percentile | Top 20th vs. lowest 20th percentile | Top 30th vs. lowest 30th percentile |
| | | *Best P-thres* | AUC | | | | |
| SCZ vs. MDD | **0.763** | 0.005 | 0.694 | 11.31 | 6.61 | 3.46 | 2.17 |
| BPD vs. MDD | **0.749** | 0.005 | 0.602 | 2.82 | 2.24 | 1.70 | 1.39 |
| MDD vs. ED | **0.78** | 0.0005 | 0.587 | 1.09 | 1.07 | 1.04 | 1.03 |
| MDD vs. ASD | 0.694 | 0.01 | 0.551 | 1.10 | 1.08 | 1.05 | 1.03 |
| MDD vs. ADHD | **0.707** | 0.005 | 0.583 | 1.38 | 1.29 | 1.18 | 1.11 |
| MDD vs. anxiety | **0.744** | 0.001 | 0.57 | 1.54 | 1.40 | 1.25 | 1.15 |
| MDD vs. OCD | **0.8** | 0.005 | 0.566 | 1.12 | 1.09 | 1.06 | 1.04 |
| SCZ vs Psychotic experiences | **0.828** | 0.005 | 0.686 | 9.19 | 5.62 | 3.11 | 2.03 |
| BPD vs Psychotic experiences | **0.801** | 0.005 | 0.584 | 2.02 | 1.73 | 1.43 | 1.25 |
| MDD vs Psychotic experiences | 0.668 | 0.00001 | 0.524 | 1.09 | 1.07 | 1.04 | 1.03 |
| SCZ vs. BPD | **0.726** | 0.005 | 0.618 | 3.29 | 2.53 | 1.84 | 1.46 |
| ADHD vs. ASD | **0.804** | 0.001 | 0.622 | 1.50 | 1.37 | 1.23 | 1.14 |

AUC: area under the ROC curve; Maximum: the maximum discriminating ability (in AUC) for the 2 traits in each comparison based on SNP-based heritability, assuming that all risk variants in the GWAS panel were found; Values of AUC above 0.7 were in bold. Best P-thres: the p-value threshold for PRS analysis leading to the highest AUC. The last four columns show the relative risk of the 1$^{st}$ disorder at the top $x^{th}$ against the lowest $x^{th}$ percentiles of PRS (derived from treating disorder A as 'case' and disorder B as 'control')

MDD vs PTSD was not listed above as the results from GWAS analysis were too weak in power so the AUC based on PRS of current GWAS data cannot be estimated. The maximum AUC (based on SNP-based $h^2$) was estimated to be 0.588.

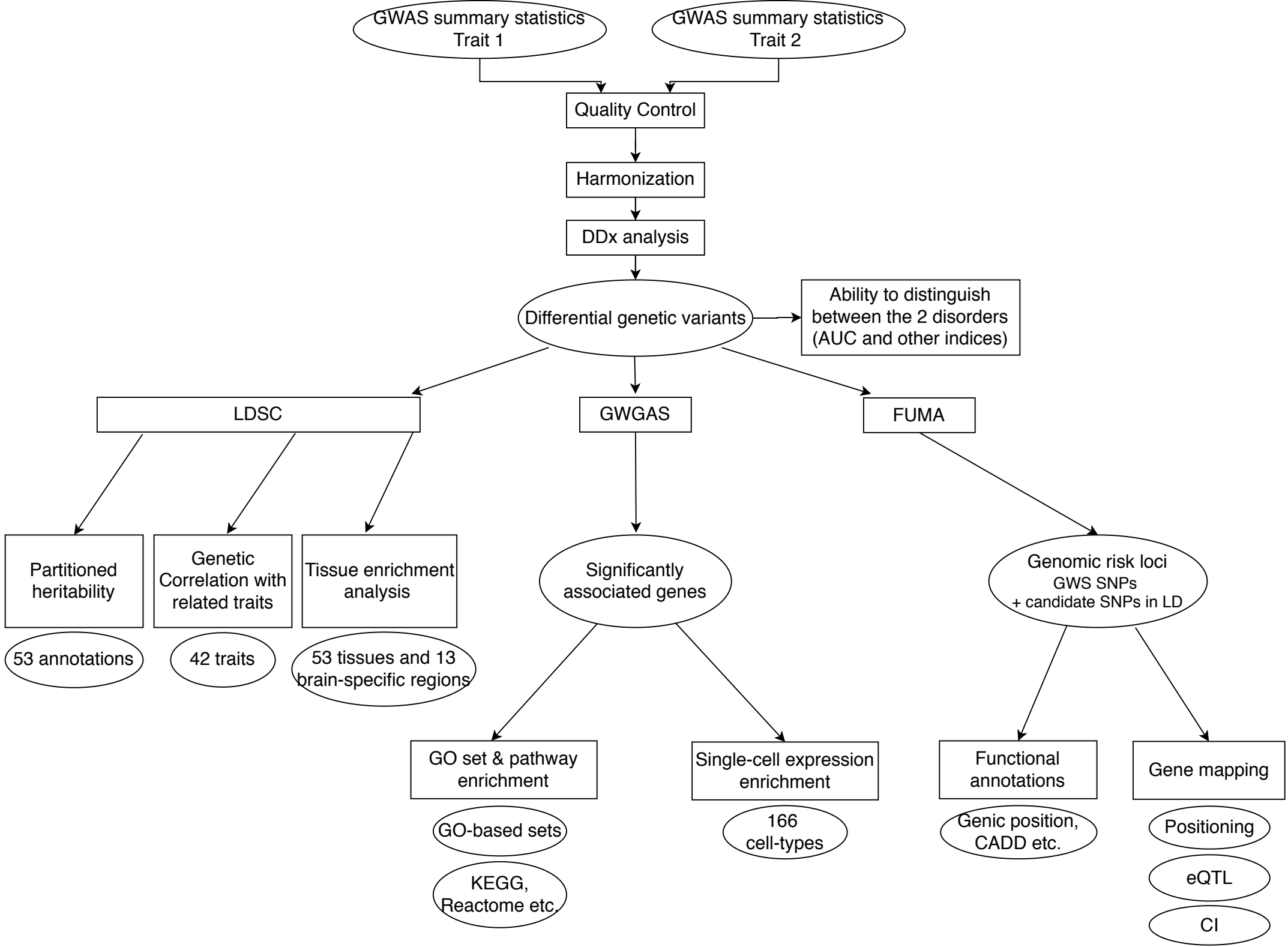

# Analysis of genetic differences between psychiatric disorders: exploring pathways and cell-types/tissues involved and ability to differentiate the disorders by polygenic scores

Shitao RAO, Liangying YIN, Yong XIANG, Hon-Cheong SO

## Supplementary Text

### Supplementary Methods

**Quality control of SNPs and processing**

For further quality control, SNPs with a low imputation quality score (INFO $R^2 < 0.6$) were excluded from further analysis. In addition, indels and duplicated SNPs were filtered. We then performed a harmonization step to keep the reference allele for signed test statistics consistent between each pair of GWAS datasets. Following that, the post–quality-control and harmonized summary statistics were utilized for investigating differential genetic variants for 26 comparisons of psychiatric disorders/traits (Table 3) using a statistical method presented below.

**Psychiatric traits/disorders included in the analysis**

As for the choice of disorders to compare, we intend to select pairs of disorders/traits which have some epidemiological associations and/or genetic overlap but are also distinct entities. For example, MDD is associated with a variety of other psychiatric disorders, such as SCZ[1], anxiety disorders[2], PTSD[3], OCD[4], alcohol, cannabis and other psychoactive substance use[5-8], ASD[9], ADHD[10], ED[11] etc., but each disorder also has its distinct clinical characteristics and aetiologies. As another example, neuroticism has been reported to be associated with other psychiatric disorders such as anxiety disorders, SCZ, MDD and AD [12-14]. In addition, we included several other pairs of comparisons such as SCZ vs BPD[15], ASD vs ADHD[16] and AD vs cannabis use[17], and they are related but also distinct psychiatric traits/diagnoses. We have included cannabis use as it is related to many psychiatric disorders[18], and at the time of analysis, it is the only trait related to substance abuse (excluding alcohol dependence) with publicly available GWAS data and reasonably large sample sizes. Ideally a more well-characterized disorder or phenotype should be studied, which will be considered in our future work. Some other comparisons also included anxiety disorders and suicide attempts[19] and psychotic experience with SCZ, BPD[20] and MDD[21, 22] due to possible clinical and/or genetic links between these traits/disorders.

*A note on the MDD GWAS sample*

For MDD which formed a major part of our comparisons, a recent GWAS meta-analysis was carried out based on 135,458 cases and 344,901 controls[23]. Excluding 23andMe data, the released GWAS summary statistics were generated from a sample set of 59,851 cases and 113,154 controls with a higher SNP-based heritability(7.8%, se 0.5%). The majority of MDD cases (45591 of 59851 cases, ~76.2%) in this sample were defined by clinical assessment or clinical records according to ICD/DSM criteria, although the UKBB sub-sample(14260/59851 cases) included some cases from self-reporting. While an updated study[24] included a larger sample, the majority (excluding 23andMe) was composed of the broad depression phenotype in the UK Biobank dataset(127552/170756 cases); the sample also showed a lower SNP-based heritability(6.0%, se=0.3%)[24]. A recent study showed that genetic studies on depression defined by minimal or 'broad' phenotyping may not be specific to MDD itself[25]. Such studies might identify non-specific genetic factors linked to other psychiatric conditions; this may defy our purpose of finding differential genetic markers between related disorders/traits.

**Functional annotations of identified differential genetic markers**

The differential genetic variants identified were further explored for their biological functions using FUMA (https://fuma.ctglab.nl/)[26]. Following the definition by FUMA, independent significant SNPs were defined as those with $p<5e-8$ and independent from each other at the default $r^2$ threshold ($r^2=0.6$). For the definition of genomic *loci*, independent significant SNPs which are correlated with each other at $r^2 \geq 0.1$ are assigned to the same risk locus. Independent significant SNPs which lie within 250 kb are also merged into one genomic risk locus. All candidate SNPs in defined risk loci that are in LD ($r^2 \geq 0.6$) with the corresponding independent significant SNPs were selected for functional annotations, mainly including combined annotation-dependent depletion (CADD) scores[27], chromatin states[28, 29], ANNOVAR categories[30] and RegulomeDB scores[31].

**Gene mapping**

SNPs were mapped to genes in FUMA using three different strategies including mapping by position, expression quantitative trait loci (eQTL), and chromatin interactions (CI). In brief, the positional method maps variants to genes based on their physical position, while the eQTL strategy maps SNPs to genes with which a significant (FDR<0.05) eQTL association exists. The third strategy (CI) maps SNPs to genes based on three-dimensional (3D) DNA-DNA interaction of the SNP and gene regions.

**Genome-wide gene-based association study (GWGAS) and tissue/cell-type enrichment analysis**

*P*-values from SNP-based analysis were utilized for GWGAS analysis in MAGMA[32]. The program aggregates statistical significance of SNPs within a gene to output a gene-based statistic. Multiple testing was corrected by the false discovery rate (FDR) approach. In our gene-based and other analyses to follow, results with FDR<0.05 were considered significant.

The biological functions of GWGAS-significant genes were further investigated *via* tissue and cell-type expression enrichment analysis using MAGMA[32] and Linkage Disequilibrium SCore regression (LDSC)[33]. In tissue enrichment analysis, MAGMA was used to test for enrichment based on over-representation of differentially expressed genes (DEGs) in each of 53 tissues in GTEx. We observed that brain regions were predominantly enriched in the above analysis; hence we focused subsequent analyses on the brain. Next we conducted an enrichment analysis *within* 13 brain regions using LDSC, based on GTEx data. This is a 'competitive' analysis restricted to the brain; the aim was to reveal enrichment within specific brain regions when compared to other regions.

Following that, all available single-cell expression datasets from human brain regions [Lateral Geniculate Nucleus (LGN), Middle Temporal Gyrus (MTG), hippocampus, cortex, prefrontal cortex, midbrain and temporal cortex] in FUMA were included for enrichment analysis to explore the specific types of contributing cells/neurons. A 2-step workflow was implemented for the enrichment analysis. The 1$^{st}$ step was carried out to identify significantly enriched cell types, which were retained for the 2$^{nd}$ step to determine independent signals within a dataset by stepwise conditional analysis (see also https://fuma.ctglab.nl/tutorial#celltype).

We also conducted pathway and gene-set enrichment analyses to explore whether these significantly associated genes were significantly enriched in biological predefined pathways or gene ontology (GO) sets based on the ConsensusPathDB database (CPDB, human) (http://consensuspathdb.org/)[34].

**SNP-based heritability and genetic correlation with related traits**
SNP-based heritability ($h^2_{snp}$) of differential genetic associations was estimated by LDSC and SumHer [35]. The former is the most widely used approach for estimating $h^2_{snp}$, and was employed as the primary estimation method here. We also performed additional analysis with SumHer, another program for $h^2_{snp}$ estimation that allows more realistic heritability models. We also conducted 'partitioned heritability' analysis to identify which functional categories of genetic variants (e.g coding, promoter, histone marks, enhancers etc.) contribute the most to differentiation of the psychiatric disorders/traits [36].

In addition to shedding light on genetic architecture and relative importance of different functional categories, heritability explained is connected to the predictive power of genetic variants [37]. In this regard, we also estimated the *maximum* 'predictive ability' (ability to differentiate the disorders in our case) that can be achieved if all variants on the GWAS panel are accounted for. Liability-scaled SNP-based heritability ($h^2_{snp}$) was calculated using LDSC, with sample and population prevalence as input. We estimated the 'sample prevalence' by the effective number of cases[38] in the two datasets. We then followed the methodology described in [37] to compute different predictive indices and graphs. Briefly, we computed the AUC under ROC curve, proportion of cases explained by those at the top $k$% of predicted risk, variance of predicted risk and the absolute risk at different percentile. The graphs included ROC curve, predictiveness curve and the probability and cumulative density function of predicted risks. The analysis on differentiating ability was performed on comparisons of selected psychiatric disorders (SCZ, BPD, ED, ASD, ADHD, anxiety disorders, PTSD, OCD) and clinical symptoms (psychotic experience) for which differential diagnosis is considered more clinically relevant.

Genetic correlations ($r_g$) between the differential genetic variations and 42 potentially related phenotypes were calculated using LDSC(http://ldsc.broadinstitute.org/centers/)[39]. Generally, $r_g$ reflects how much the non-shared or unique genetic component of the 1$^{st}$ disorder is genetically correlated with a specific trait, when compared to the 2$^{nd}$ disorder in the pair. The rationale of this analysis is that when we consider two disorders as different, it is important to see whether they are associated with *different comorbid disorders/traits*. This distinction is important to help understand the different prognosis or aetiology of different disorders. For example, SCZ is generally associated with more prominent cognitive deficits than BPD. The above analysis may help to highlight such differences.

A set of 42 GWAS summary statistics were obtained from the LD-Hub[39] and grouped into nine categories of traits including neurological diseases, personality traits, sleeping, cognitive, education, brain volume, psychiatric disorders, cardiometabolic traits and aging.

**Ability of polygenic risk scores (PRS) from existing GWAS data to differentiate disorders**

For selected traits for which differential diagnosis (DDx) are more clinically relevant, we performed another analysis to evaluate the ability of polygenic risk scores (PRS) from *existing* GWAS data to distinguish psychiatric disorders. The PRS was based on a case-control study of the corresponding disorders (disorder A as 'case' and disorder B as 'control'). Note that unlike above, we are *not* focusing on the *maximum* predictive power achievable from all common variants.

An empirical Bayes approach has been proposed to recover the underlying effect sizes and could be used to forecast predictive ability of PRS, based on summary statistics alone[40]. The method has been verified in simulations and real data applications[40]. Eighteen subsets of genetic variants based on a series of *P*-value thresholds ($10^{-5}$, $10^{-4}$, $5\times10^{-4}$, $10^{-3}$, $5\times10^{-3}$, 0.01, 0.03, 0.05, 0.1, 0.2, 0.3, 0.4, 0.5, 0.6, 0.7, 0.8, 0.9, and 1) were used to construct PRS.

**Additional details on the simulation model**

More specifically, we simulated standard normal variables $z_i \sim N(0,1)$, and set mean effect size $\mu = \sqrt{\frac{h^2}{N_{snp}}}$. The actual effect size for $SNP_i$ was set at $\beta_i = \mu * z_i$. The total liability $y$ equals the sum of effects from each SNP plus a residual ($e$), i.e. $y = \sum_i \beta_i x_i + e$; the total variance of $y$ was set to one. Following the liability threshold model, subjects with total liability exceeding a certain threshold [$= \Phi^{-1}(K)$, where $K$ is the disease prevalence] are regarded as having the trait/disease. The non-shared genetic covariance between the two traits was set to 0.1.

**Supplementary Results**

Here we provide a more detailed description of the results.

**MDD against psychiatric disorders/outcomes**

In this part, we compared MDD with 12 different psychiatric disorders/outcomes, including SCZ, BPD, ED, ASD, ADHD, anxiety disorder, insomnia, alcohol dependence, ever used cannabis, SA, PTSD and OCD (Table 3). Totally 69 genomic risk loci were identified from the 12 pairs of comparisons(Table 3). Please refer to Table S1 to S12 for detailed results.

*MDD against SCZ*

Among the 12 pairs of comparisons, comparison of MDD and SCZ generate the largest number of genome-wide significant SNPs (2,312 SNPs, Table 3; sub-table 1 in Supplementary Table 1 (Table S1.1)) which belong to 37 genomic risk loci (Table S1.2). Although most of the candidate variants were located in intergenic and intronic regions (81% of variants) (Table S1.3), 65 SNPs were located in exons, including 32 nonsynonymous variants. Heritability enrichment analyses of 53 functional annotation categories indicated that the heritability of SNPs was not only enriched in intronic and conserved regions [Table S1.4; $P<1.28E-04$], but also in coding regions including transcription start site (TSS) ($P = 1.78E-04$). Besides, regulatory categories such as methylation and acetylation marks were found to be enriched among significant variants [$P < 1.57E-04$].

The three gene-mapping strategies (positional, eQTL and CI mapping) generated a set of 524 unique genes, 94 of which were implicated by all three methods (Table S1.5). Additionally, GWGAS analysis identified 953 significant genes (Table 3; Table S1.6). Taken together, 64 genes were implicated by all four strategies. Among them, *CACNA1C* was predicted to have a very high probability of loss of function mutation intolerance (pLI score=1; Table S1.5). Genes differentiating MDD and SCZ were mainly enriched in the cortex, the anterior cingulate cortex (BA24), and the frontal cortex (BA9) regions (Table S1.8;FDR<6.0E-04). Cell-type enrichment analysis suggested strong associations with several kinds of neurons in the cortex and prefrontal cortex (Table S1.9). Moreover, this analysis also identified associations with neurons in the midbrain, hippocampus, and lateral geniculate nucleus(LGN) regions(Table S1.9). Conditional analyses suggested neurons in the cortex, GABAergic neurons in the midbrain, and pyramidal neurons in the hippocampus as *independent* contributing neurons (after controlling for other cell types) (Table S1.10).

In gene-set enrichment analysis(GEA), the 953 GWGAS significant genes were enriched in a number of biological GO sets, including generation of neurons, regulation of nervous system development and central nervous system neuron differentiation [Table S1.11; FDR< 5.88E-03]. Other enriched pathways include neuronal system, alcoholism and brain-derived neurotrophic factor (BDNF) signalling pathway [Table S1.12; FDR< 1.86E-02]. In genetic correlation analysis, SCZ was defined as 'case' and MDD as (pseudo-)'control'. Note that a positive genetic correlation indicates that the 'case' disorder is more positively associated with the studied trait genetically than the (pseudo-)'control' disorder, and vice versa. For example, we observed inverse genetic correlations(rg) with insomnia, neuroticism, coronary artery disease (CAD) and mean hippocampal volume, among others. This suggested that MDD has stronger positive genetic correlations with the above traits/disorders compared to SCZ. Findings of this type may shed light on different patterns of comorbidities, but may also be clinically informative. For instance, the significant inverse rg with CAD suggested that compared to SCZ patients, MDD patients may be more genetically predisposed to CAD.

*MDD against BPD, ED, ASD, ADHD, Anxiety disorder, Insomnia, Alcohol dependence and Cannabis use*

In these 8 pairs of comparisons, we identified 32 differential genomic loci (Table 3; detailed in Table S2-S9). The comparison between MDD and BPD revealed the largest number of significant genes based on GWGAS (174 genes; Table 3; Table S2.6). Here we just briefly highlight the comparisons of MDD with a few disorders (BPD, ADHD and anxiety disorders) which yielded the largest number of significant genes in GWGAS.

In the comparison between MDD and BPD, we found 4 significant risk loci, in which the strongest signal rs17751061 was found to be highly pathogenic likely influencing the function of *SUGP1* (Table S2.2, CADD = 35). Another nonsynonymous variant, rs17420378, was located in exon 8 of *STK4* (risk loci no. 4) with high predicted pathogenicity(Table S2.2, CADD=22.7). Besides, *STK4* was implicated by all four gene-mapping strategies(Table S2.5 and S2.6). GEA revealed 174 GWGAS-significant genes, which were enriched in 17 gene ontology (GO) sets[Table S2.10; $P<0.01$], such as transitional metal ion binding and pre-mRNA binding. Our pathway enrichment analysis indicated that differential genetic associations between MDD and BPD were involved in neural cell adhesion molecule (NCAM) signalling for neurite outgrowth, amphetamine addiction, serotonergic and glutamatergic synapse [Table S2.11;FDR = 4.88E-02 for all four pathways], among other pathways.  The differential variants were enriched in brain regions including the cerebellum, cortex, and frontal cortex (BA9) [Table S2.7; FDR < 4.03E-02]. Enrichment analysis *within* brain regions suggested that the cortex and frontal cortex were the most enriched compared to others[Table S2.8; FDR<5.31E-04]. The most significantly enriched cell type was GABAergic neurons from LGN. Interestingly, the differential variants (BPD vs MDD) were found to have positive correlations with childhood IQ and a higher level of education [Table S2.12;FDR<3.26E-02], but negative correlations with insomnia and CAD [FDR< 4.92E-03].

In the comparison between MDD and anxiety disorders, five genetic loci were found (Table 3), one of which involved the extended MHC (xMHC) region [41]. The top mapped genes involved a set of genes in the xMHC region, and 3 other genes from other chromosomes (*LRFN5, PTPN1, FAM65C*) (Table 4). We note that due to the complex LD structure and high gene density in this region, it may be relatively difficult to identify the true casual gene/variant. GWGAS revealed 106 significant genes(Table S6.6). The most enriched tissues contributing to differential associations included the nucleus accumbens, frontal cortex and cerebellar hemisphere (Table S6.7); the most enriched cell types included GABAergic neurons from hippocampus, midbrain and temporal cortex, among others (Table S6.9).

In the comparison of MDD with ADHD, 167 significant differential genetic variants were identified which formed 5 genetic loci (Table S5.1 and S5.2). Four genes were mapped by all 3 gene-mapping strategies, including *KDM4A, SLC6A9, TMEM161B and CDH8* (Table S5.5 and S5.6). Altogether 82 genes were significant in gene-based test, and pathway and GSEA shed light on pathways such as those related to DNA methylation (Table S5.10 and S5.11). The most enriched cell types included GABAergic neurons in the midbrain and prefrontal cortex, as well as dopaminergic neurons in the midbrain (Table S5.9).

**MDD against depression-related traits**

In this section, we tried to identify differential genetic variants from three pairs of comparisons between MDD and three depression-related phenotypes (probable recurrent severe depression, seen GP for anxiety/depression and longest period of feeling low/depressed). Ten risk loci were identified (Table 3, Table S13 to S15).

*MDD against depression defined in UKBB*

First we compared MDD (from PGC; majority clinically defined) against probable recurrent depression (severe)[ProbDep]. We identified 4 risk loci (Table 3;Table S13.2), including one in the xMHC region. Gene-based test revealed 110 significant genes. Tissue enrichment analysis highlighted the cerebellar hemisphere, nucleus accumbens and frontal cortex as the most enriched regions. Cell-type enrichment analysis suggested that the significant genes were associated with GABAergic, dopaminergic and other types of neurons in LGN, middle temporal gyrus (MTG), hippocampus, midbrain and cortex regions(Table S13.9). Pathway analysis mainly highlighted those related to DNA methylation and histone modification contributed by histone genes in the xMHC region; other top pathways included Rett syndrome causing genes and axon guidance pathway. GO sets enriched included regulation of long-term neuronal synaptic plasticity, central nervous system neuron development and differentiation(Table S13.11 and S13.12). Genetic correlation analysis showed that MDD-PGC was more positively genetically correlated with most other psychiatric disorders (e.g. SCZ, BPD, ASD, ADHD) as well as CAD when compared with ProbDep(Table S13.13). We then compared MDD against seen GP for nerves/anxiety/depression(GPDep). The significant variants mapped into 3 loci, 2 of which were also observed in the above analysis (including one in the xMHC region). Gene-based analysis revealed 72 genes; we observed an overlap of 69 genes with the previous analysis with ProbDep, although the latter analysis identified 110 significant genes. Other results of comparison between MDD and GPDep are shown in detail in Tables S14.

*MDD against duration of longest period of feeling low/depressed (top quintile as case)*

For this comparison, functional annotation of 133 candidate SNPs formed 3 genetic risk loci, among which the *GRIK2* gene was also mapped by the three gene-mapping methods (positional, eQTL, CI mapping; Table S15.5). The gene codes the Glutamate Ionotropic Receptor Kainate Type Subunit 2, suggesting glutamatergic transmission may be one factor with differential associations between susceptibility to depression and severity (as reflected by duration) of illness. Possibly due to limited sample size, tissue and cell-type enrichment analysis did not reveal significant results.

**Neuroticism against SCZ/MDD/Anxiety disorder/alcohol dependence**

In this part, five sets of GWAS summary statistics were employed which formed four pairs of comparisons (neuroticism against anxiety disorder, SCZ, MDD and alcohol dependence; Table 2). The choice is based on relatively high association of neuroticism with these disorders [12-14]. We identified 1,294 genomic risk loci from the four comparisons (Table 3). For space limits, we highlight the results of neuroticism vs MDD only (Table S16). Please refer to Tables S17-S19 for detailed findings of other comparisons.

In the comparison of neuroticism against MDD, 20 risk loci were identified(Table S16.1 and S16.2). Functional annotations of 5,573 candidate SNPs in these loci highlighted a number of genes, among which *CRHR1, MAPT, WNT3* and *KANSL1* were mapped by all 3 gene-mapping methods and MAGMA (Table S16.5 and S16.6). They all belong to a risk locus on chr 17 but the exact causal gene(s) may require clarification in further studies. Tissue enrichment analysis observed enriched signals in most brain regions(Table S16.7); within-brain comparison showed that the cortex, frontal cortex, anterior cingulate cortex and nucleus accumbens were the most enriched [Table S16.8; FDR<1.40E-02]. Cell-type enrichment analysis of the GWGAS-significant genes suggested that they were mainly enriched in the LGN region (*P* (FDR within one dataset)<4.40E-02; Table S18.9). GO set enrichment analysis revealed that axon extension, CNS neuron differentiation and regulation of neuron death may be involved (Table S16.10).

**Psychotic experiences against SCZ/BPD/MDD**
Here we identified 10 and 2 genomic risk loci from comparison of psychotic experiences against SCZ and BPD respectively, but not from psychotic experiences against MDD (Table 3; Table S20-S22).

*Psychotic experiences against SCZ*

In this comparison, functional annotation of 1,749 candidate SNPs revealed 10 genomic risk loci, covering 82 genes (Table S20.2). Altogether 68 genes were mapped by all three gene-mapping strategies, over half of which (35/68) was also indicated by GWGAS analysis (Table S20.5 and S20.6). The most implicated brain regions were the cortex, frontal cortex(BA9) and anterior cingulate cortex (Table S20.8). Cell-type enrichment analysis also observed significant signals in cortex and prefrontal cortex as the top two findings (Table S20.9). Enriched pathways or gene-sets included anterograde trans-synaptic signalling, synaptic vesicle exocytosis/ localization and synaptic adhesion-like molecules (Table S20.11 and S20.12).

*Psychotic experiences and BPD*

Functional annotation analysis highlighted several genes (*SUGP1, GATAD2A and CILP2*) harbouring SNPs with very high CADD scores (CADD score> 13.31, Table S21.2). The 3 gene-mapping strategies mapped variants to 10 genes(Table S21.5), all of which were also identified by GWGAS(Table S21.6). Tissue enrichment analysis suggested enrichment in the cortex, cerebellar hemisphere, frontal cortex, anterior cingulate cortex and cerebellum regions (FDR<2.99E-02; Table S21.7 and S21.8). Further analysis highlighted cocaine and amphetamine addiction, PTEN and EGF signalling as top pathways (Table S21.11).

*Psychotic experiences and MDD*

Although the comparison did not generate any significant differential variants, a few genes were highlighted via functional annotations of candidate SNPs (Table S22.2). All of the three gene-mapping strategies linked variants to 9 protein-coding genes (*TMEFF2, SLC30A9, BEND4, PRLR, EPM2A, FBXO30, SHPRH, GRM1, ANK3*; Table S22.5). Genetic correlation analysis suggested that MDD showed stronger rg with SCZ or BPD compared to psychotic experiences (Table S22.10).

**Other pairs of comparisons**

We also applied the proposed methods to the other four clinically relevant comparisons, including SCZ against BPD, ADHD against ASD, alcohol dependence against ever used cannabis and anxiety disorder against SA (Table 3). We identified 3, 7, 2 and 1 genomic risk loci from each of the comparison respectively (Table 3). Here we highlight the results from SCZ vs BPD and ADHD vs. ASD as examples (please refer to Tables S23-S26 for details).

*SCZ vs BPD*

Our analytic results based on GWAS summary data showed almost perfect genetic correlation with those obtained by comparing BPD and SCZ using *individual* genotype data [42] ($r_g$=1.054, se=0.025).

As for the actual results, we observed three significant loci in the comparison of SCZ and BPD (Table 3). GWGAS highlighted 144 significant genes which were enriched for brain regions [Table S23.7; FDR<1.87E-02]. The frontal cortex and anterior cingulate cortex were the most enriched regions compared to other brain regions [Table S23.8; FDR< 2.55E-02]. Furthermore, cell-type enrichment analysis identified an enrichment signal in three different types of neurons in the midbrain region [Table S23.9; *P*<4.01E-03], which could withstand multiple testing correction within the corresponding dataset(FDR=3.34E-02). The enriched pathways included the inositol metabolism pathway and those related to cellular senescence [Table S23.11; FDR = 2.51E-02].

*ADHD vs ASD*

In the comparison between ADHD and ASD, 7 risk loci were found (Table S24.2). Seven genes (*KDM4A, ERI1, SOX7, PINX1, XKR6, MTMR9 and SEMA6D*; Table S24.5 and S24.6) were highlighted by all three gene-mapping methods and GWGAS (Table S24.5 and S24.6). We observed that ADHD may have stronger positive rg with CAD and insomnia compared to ASD; however, the reverse was observed for years of education, parental age at death and intra-cranial volume (Table S24.12).